\documentclass[twocolumn]{aastex63}
\usepackage{graphicx}
\usepackage{hyperref}
\usepackage{latexsym}
\usepackage{natbib}
\usepackage{amssymb}
\usepackage{amsmath}
\usepackage{url}
\usepackage{soul}
\usepackage[normalem]{ulem}
\accepted{for publication in AJ on July 23, 2020}

\newcommand\ER{\hbox{R$_\oplus$ }}
\definecolor{aqua}{rgb}{0.0,0.67,1.0}
\hypersetup{citecolor=aqua, linkcolor=aqua}

\begin{document}

\title{$\pi$ Earth: a 3.14-day  Earth-sized Planet from $\textit{K2}$'s Kitchen Served Warm by the SPECULOOS Team}

\author[0000-0002-8052-3893]{Prajwal Niraula}
\affiliation{Department of Earth, Atmospheric and Planetary Sciences, MIT, 77 Massachusetts Avenue, Cambridge, MA 02139, USA}

\author[0000-0003-2415-2191]{Julien de Wit}
\affiliation{Department of Earth, Atmospheric and Planetary Sciences, MIT, 77 Massachusetts Avenue, Cambridge, MA 02139, USA}

\author[0000-0002-3627-1676]{Benjamin V. Rackham}
\altaffiliation{51 Pegasi b Fellow}
\affiliation{Department of Earth, Atmospheric and Planetary Sciences, MIT, 77 Massachusetts Avenue, Cambridge, MA 02139, USA}

\author{Elsa Ducrot}
\affiliation{Astrobiology Research Unit, University of Li\`ege, All\'ee du 6 Ao\^ut, 19, 4000 Li\`ege (Sart-Tilman), Belgium}

\author[0000-0001-9892-2406]{Artem Burdanov}
\affiliation{Department of Earth, Atmospheric and Planetary Sciences, MIT, 77 Massachusetts Avenue, Cambridge, MA 02139, USA}

\author{Ian J. M. Crossfield}
\affiliation{Kansas University Department of Physics and Astronomy, 1082 Malott, 1251 Wescoe Hall Dr. Lawrence, KS 66045}

\author[0000-0003-2144-4316]{Val\'erie Van Grootel}
\affiliation{Space Sciences, Technologies and Astrophysics Research (STAR) Institute, University of Li\`ege, All\'ee du 6 Ao\^ut 19C, B-4000 Li\'ege, Belgium}

\author{Catriona Murray}
\affiliation{Cavendish Laboratory, J J Thomson Avenue, Cambridge, CB3 0HE, UK}

\author{Lionel J. Garcia}
\affiliation{Astrobiology Research Unit, University of Li\`ege, All\'ee du 6 Ao\^ut, 19, 4000 Li\`ege (Sart-Tilman), Belgium}

\author{Roi Alonso}
\affiliation{Instituto de Astrof\'\i sica de Canarias, E-38205 La Laguna, Tenerife, Spain}
\affiliation{Dpto. de Astrof\'isica, Universidad de La Laguna, 38206 La Laguna, Tenerife, Spain}

\author[0000-0001-7708-2364]{Corey Beard}
\affiliation{Department of Physics \& Astronomy, The University of California, Irvine, 4129 Frederick Reines Hall, Irvine, CA 92697, USA}

\author{Yilen G\'omez Maqueo Chew}
\affiliation{Instituto de Astronom\'ia, Universidad Nacional Aut\'onoma de M\'exico, Ciudad Universitaria, CDMX, C.P. 04510, Mexico}

\author{Laetitia Delrez}
\affiliation{Astrobiology Research Unit, University of Li\`ege, All\'ee du 6 Ao\^ut, 19, 4000 Li\`ege (Sart-Tilman), Belgium}
\affiliation{Space Sciences, Technologies and Astrophysics Research (STAR) Institute, University of Li\`ege, All\'ee du 6 Ao\^ut 19C, B-4000 Li\'ege, Belgium}
\affiliation{Observatoire de l’Universit\'e de Gen\`eve, Chemin des Maillettes 51, Versoix, CH-1290, Switzerland}

\author[0000-0002-9355-5165]{Brice-Olivier Demory}
\affiliation{University of Bern, Center for Space and Habitability, Gesellschaftsstrasse 6, 3012 Bern, Switzerland}

\author[0000-0003-3504-5316]{Benjamin J. Fulton}
\affiliation{NASA Exoplanet Science Institute/Caltech-IPAC, Pasadena, CA, USA}

\author[0000-0003-1462-7739]{Micha\"el Gillon}
\affiliation{Astrobiology Research Unit, University of Li\`ege, All\'ee du 6 Ao\^ut, 19, 4000 Li\`ege (Sart-Tilman), Belgium}

\author[0000-0002-3164-9086]{Maximilian N. G\"unther}
\altaffiliation{Juan Carlos Torres Fellow}
\affiliation{Department of Physics, and Kavli Institute for Astrophysics and Space Research, Massachusetts Institute of Technology, Cambridge, MA 02139, USA}

\author[0000-0002-0531-1073]{Andrew W. Howard}
\affiliation{California Institute of Technology, Pasadena, California, USA}

\author[0000-0002-0531-1073]{Howard Issacson}
\affiliation{Department of Astronomy, University of California Berkeley, Berkeley, CA 94720, USA}

\author[0000-0001-8923-488X]{Emmanu\"el Jehin}
\affiliation{Space Sciences, Technologies and Astrophysics Research (STAR) Institute, University of Li\`ege, All\'ee du 6 Ao\^ut 19C, B-4000 Li\'ege, Belgium}

\author[0000-0002-5220-609X]{Peter P. Pedersen}
\affiliation{Cavendish Laboratory, J J Thomson Avenue, Cambridge, CB3 0HE, UK}

\author{Francisco J. Pozuelos}
\affiliation{Astrobiology Research Unit, University of Li\`ege, All\'ee du 6 Ao\^ut, 19, 4000 Li\`ege (Sart-Tilman), Belgium}
\affiliation{Space Sciences, Technologies and Astrophysics Research (STAR) Institute, University of Li\`ege, All\'ee du 6 Ao\^ut 19C, B-4000 Li\'ege, Belgium}

\author{Didier Queloz}
\affiliation{Cavendish Laboratory, J J Thomson Avenue, Cambridge, CB3 0HE, UK}

\author{Rafael Rebolo-L\'opez}
\affiliation{Instituto de Astrof\'\i sica de Canarias, E-38205 La Laguna, Tenerife, Spain}
\affiliation{Dpto. de Astrof\'isica, Universidad de La Laguna, 38206 La Laguna, Tenerife, Spain}

\author[0000-0001-8102-3033]{Sairam Lalitha}
\affiliation{School of Physics \& Astronomy, University of Birmingham, Edgbaston, Birmingham B15 2TT, United Kingdom}

\author{Daniel Sebastian}
\affiliation{Astrobiology Research Unit, University of Li\`ege, All\'ee du 6 Ao\^ut, 19, 4000 Li\`ege (Sart-Tilman), Belgium}

\author{Samantha Thompson}
\affiliation{Cavendish Laboratory, J J Thomson Avenue, Cambridge, CB3 0HE, UK}

\author[0000-0002-5510-8751]{Amaury H.M.J. Triaud}
\affiliation{School of Physics \& Astronomy, University of Birmingham, Edgbaston, Birmingham B15 2TT, United Kingdom}

\begin{abstract} 
We report on the discovery of a  transiting Earth-sized (0.95$R_\oplus$) planet around an M3.5 dwarf star at 57$\,$pc, K2-315b. The planet has a period of $\sim$3.14 days, i.e. ${\sim}\pi$, with an instellation of 7.45$\,$S$_{\oplus}$. The detection was made using publicly available data from $\textit{K2}$'s Campaign 15. We observed three additional transits with SPECULOOS Southern and Northern Observatories, and a stellar spectrum from Keck/HIRES, which allowed us to validate the planetary nature of the signal. The confirmed planet is well suited for comparative terrestrial exoplanetology. While exoplanets transiting ultracool dwarfs present the best opportunity for atmospheric studies of terrestrial exoplanets with the $\textit{James Webb Space Telescope}$, those orbiting mid-M dwarfs within 100$\,$pc such as K2-315b will become increasingly accessible with the next generation of observatories. \end{abstract}


\section{INTRODUCTION} 
\label{sec:intro}

\begin{figure*}[ht!]
\centering
\includegraphics[width=0.95\textwidth ]{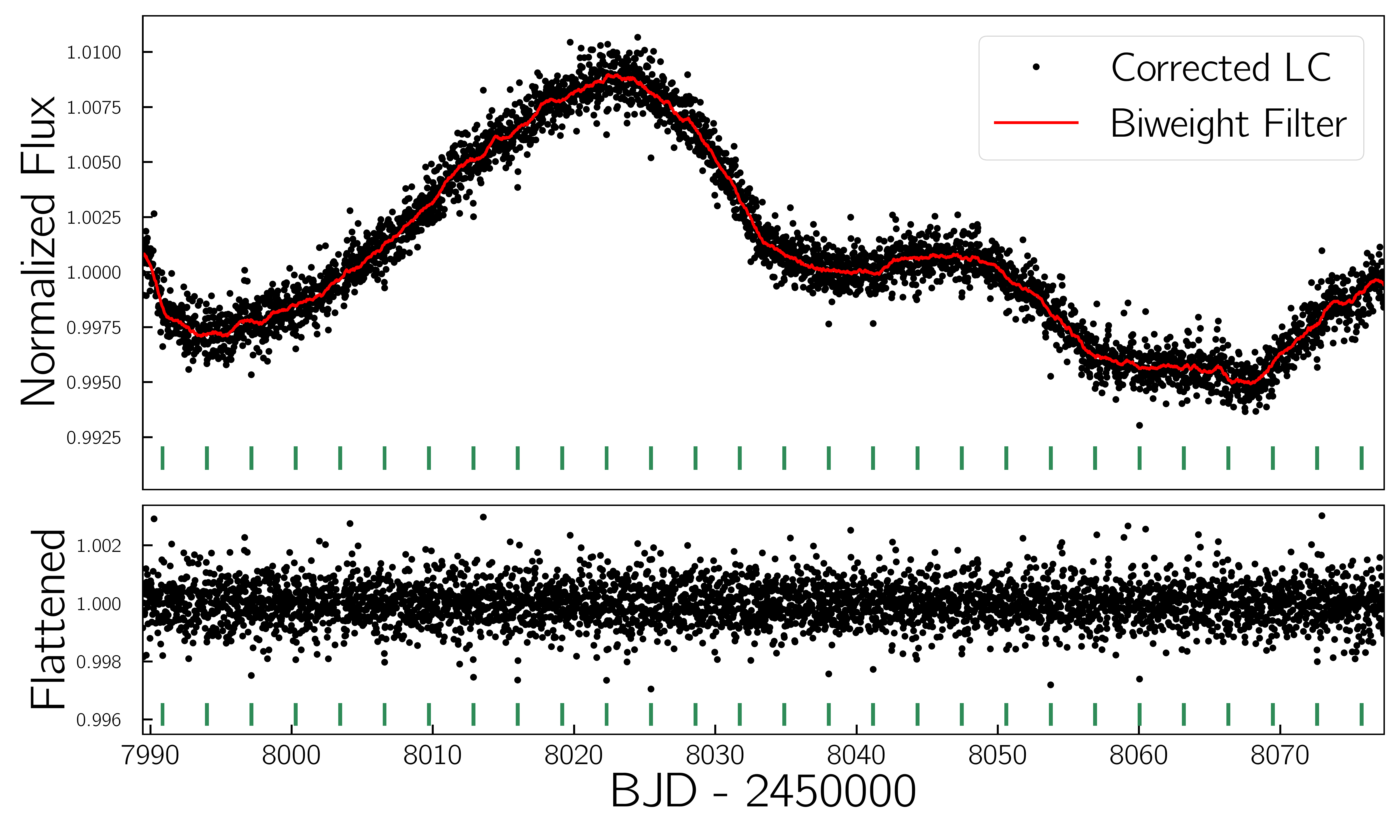} 
\figcaption{\label{fig:K2DetrendedFig}\textbf{Upper Panel:} Normalized curve using detrended light curve from \texttt{everest} pipeline of EPIC~249631677. The transits are shallow and thus not obvious. Their periodic locations are marked by green lines. The red line represents 0.75 days biweight filter used to model out the trend in the light curve potentially due to systematics and rotational modulation of the star. \textbf{Lower Panel:} Flattened light curve used for the transit fitting, and subsequent analysis in the paper.}
\end{figure*}

The redesigned \textit{Kepler} mission, \textit{K2} \citep{howell2014}, has been a success by adding almost 400 confirmed planets to the 2,348 discovered by the original mission\footnote{\href{https://exoplanetarchive.ipac.caltech.edu}{https://exoplanetarchive.ipac.caltech.edu}}. Building upon \textit{Kepler} \citep{borucki2010}, \textit{K2} expanded the search of planets around brighter stars, covered a wider region of sky along the ecliptic, and studied a variety of astronomical objects . Together, these endeavors have revolutionized the field of exoplanetary science by quadrupling the number of exoplanets known at the time, while \textit{K2} in particular has led to exciting discoveries, such as disintegrating planetesimals around the white dwarf WD-1145 \citep{vanderburg2015}, multi-planet systems around bright stars like GJ 9827 (K2-135) \citep{niraula2017, rodriguez2018}, and resonant chains of planets like the K2-138 system with five planets \citep{christiansen2018}. 

Space-based platforms such as \textit{Kepler} can provide high-quality continuous monitoring of targets above the Earth's atmosphere. The simultaneous photometric monitoring of tens of thousands of stars enables finding rare configurations (e.g., WD-1145) and answering science questions regarding planetary populations that are more statistical in nature such as how unique our own Solar System is, or what are the most common type of planets \citep[e.g.][]{fressin2013, fulton2017}.

Ground-based facilities, on the other hand, often detect fewer planets while operating at a lower cost. These planets frequently exhibit larger signal-to-noise ratios (SNRs) in various metrics (e.g., transmission), thereby allowing for these planets to be characterized further. One such example is the TRAPPIST-1 planetary system \citep{gillon2016, gillon2017}, discovered by the TRAPPIST Ultra Cool Dwarf Transiting Survey, a prototype survey for the SPECULOOS Survey \citep{gillon2013}. The goal of the SPECULOOS Survey is to explore the nearest ultracool dwarfs ($T_\mathrm{eff}<$3000\,K) for transits of rocky planets \citep{burdanov2018, delrez2018, jehin2018, Sebastian2020}. Although few systems are expected \citep{delrez2018,Sebastian2020}, their impact on the field will be significant as they should provide most of the temperate Earth-sized exoplanets amenable for atmospheric studies with the next generation of observatories such as JWST \citep[e.g.][]{gillon2020}.  

\begin{figure*}[ht!]
\centering
\includegraphics[trim={5cm 0 5cm 1cm},clip,width=0.95\textwidth ]{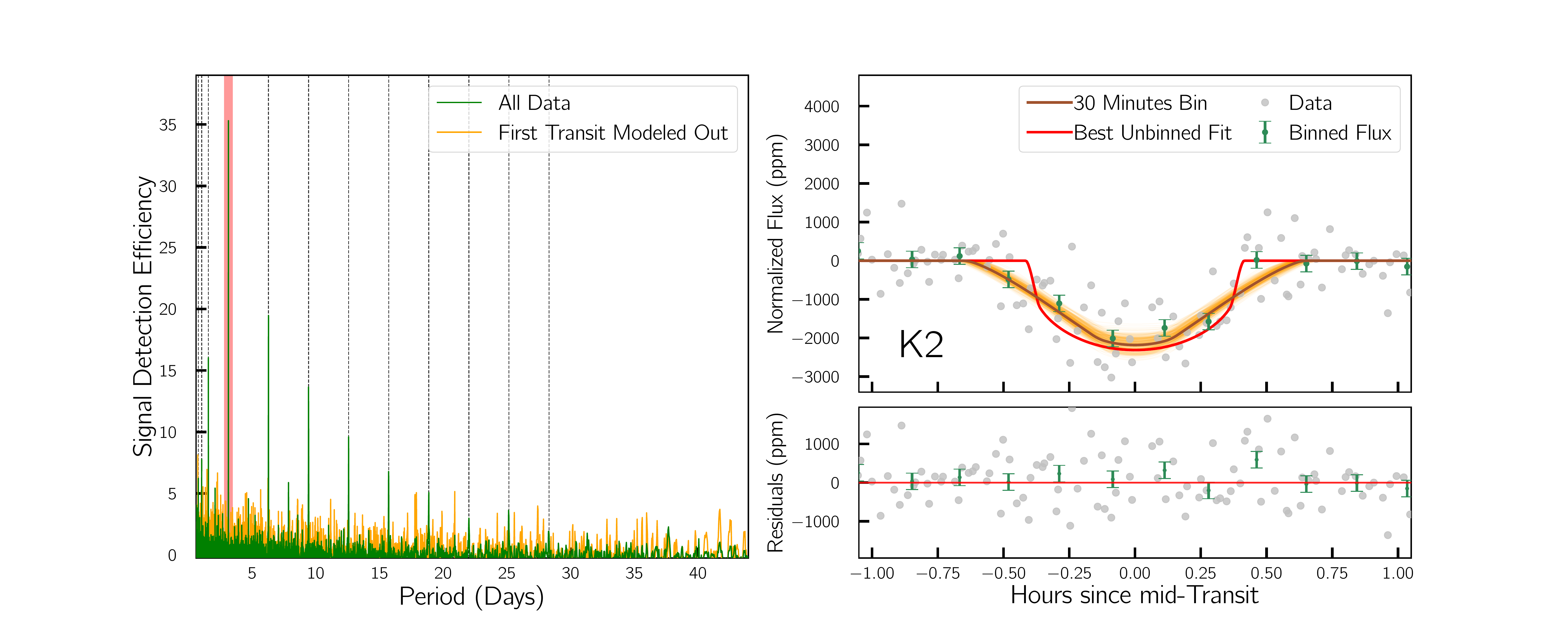} 
\figcaption{\label{fig:transitandperiodgram}\textbf{Left:} SDE obtained from TLS showing the strongest peak at $\sim$3.14 days marked in red and its aliases marked with black dotted lines. No significant additional peaks were observed once the first signal was modeled out.  \textbf{Right:} Best-fit transit model for \textit{K2} data is shown in red. The brown line is the model taking into account the integration time of 29.4 minutes for \textit{K2}. The orange lines illustrate 350 random models drawn from the posterior distributions of the fitted parameters.}
\end{figure*}

\begin{figure}[ht!]
\centering
\includegraphics[trim={0 8.5cm 0 8.5cm},clip, width=0.45\textwidth ]{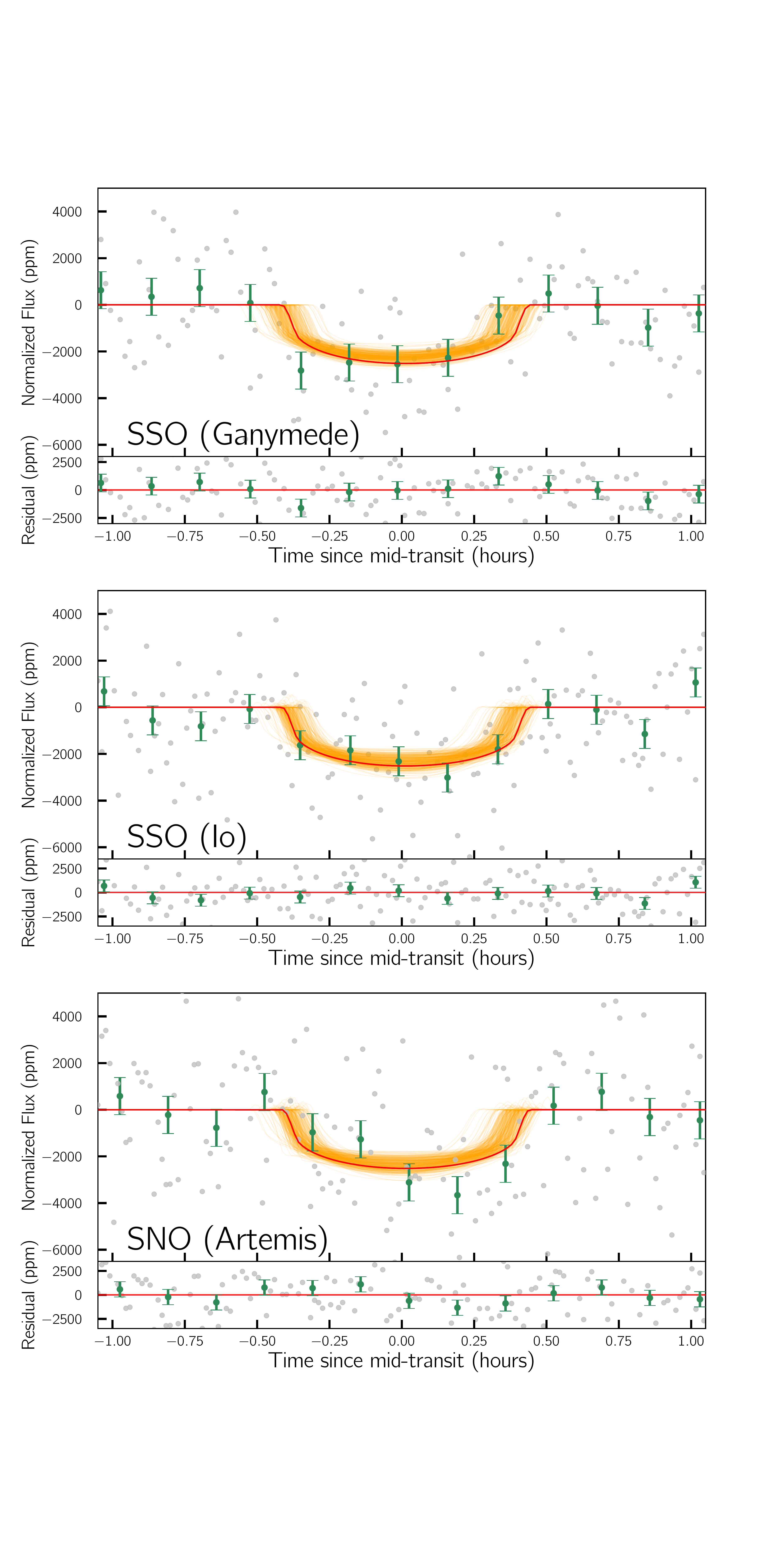} 
\figcaption{\label{fig:GNDObservation} \textbf{Top:} First ground-based observation of K2-315b from Ganymede, SSO on UT 25 February 2020  at airmass of 1.03. \textbf{Middle:} Second ground-based observation by Io, SSO on UT 18 March 2020 at airmass of 1.01. \textbf{Bottom:} Third ground-based observation by Artemis, SNO on UT 18 May 2020 at airmass of 1.77. The best-fit model, obtained from simultaneous fitting of \textit{K2} and SPECULOOS data, is shown in red with 350 randomly selected models from MCMC posteriors shown in  orange. The silver points are the detrended flux using second-order polynomials in airmass and FWHM. The green points corresponds to flux bins of 10 minutes. }
\end{figure}

Beyond the SPECULOOS Survey, which monitors nearby late-M dwarfs for terrestrial planets, the SPECULOOS telescopes have been used to study the planetary population around mid- and late-M dwarfs. In that context, SPECULOOS facilities have been involved in following up and validating planetary candidates, notably from \textit{TESS} \citep{guenther2019, kostov2019, quinn2019}. Next to confirming planetary candidates that cross detection thresholds, we have started to investigate weaker signals. For this work, we revisited  \textit{K2} data, a mission which ended only in 2019. We reanalyzed the light curves of stars with $T_\mathrm{eff} < 3500$\,K, a Kepler magnitude $< 15$, and a $\log g > 4.5$. While these criteria were motivated particularly to look for planets around ultra-cool dwarfs, they were relaxed in order to allow room for errors in the stellar properties and improve completeness of the analysis. Among the 1,213 stars fitting these criteria, EPIC~249631677 presented the strongest periodic transit-like signal. 

In this paper, we report the discovery of an Earth-sized \textit{K2} planet in a close-in orbit around EPIC 249631677. The paper is structured as follow; observations (Section~\ref{sec:obs}), analysis and  validation (Section \ref{sec:analysis}),  and the discussion in regards to future prospects for characterization (Section~\ref{sec:future}).

\section{OBSERVATIONS}
\label{sec:obs}

\subsection{A Candidate in Archival \textit{K2} Data}
EPIC~249631677 was observed by \textit{K2} in Campaign 15 from 2017-08-23 22:18:11 UTC to 2017-11-19 22:58:27 UTC continuously for about 90 days as part of program GO~15005 (PI: I. Crossfield). The pointing was maintained by using two functioning reaction wheels, while the telescope drifted slowly in the third axis due to radiation pressure from the Sun, which was corrected periodically by thruster firing \citep{howell2014}. As a consequence of such a modus operandi, uncorrected \textit{K2} light curves can show saw-tooth structures. 

Many pipelines have been built to correct for such systematics. Two popular detrending algorithms for \textit{K2} light curves are \texttt{K2SFF} \citep{vanderburg2014} and \texttt{everest} \citep{everest2016}. These pipelines have helped to achieve precision comparable to that of \textit{Kepler} by correcting for systematics caused by intra-pixel and inter-pixel variations. In the case of EPIC 249631677b,  the standard deviation of the flattened light curve for \texttt{K2SFF} was  observed to be 1230\,ppm, compared to 685\,ppm for \texttt{everest}. Considering this,  we use the light curve from the \texttt{everest} pipeline, available from the Mikulski Archive for Space Telescopes, throughout this analysis.  We use a biweight filter with a window of 0.75 days, as implemented in \texttt{w{\={o}}tan} \citep{hippke2019b}, to generate the flattened light curve for further analysis, and use only data with quality factor of 0. This light curve can be seen in \autoref{fig:K2DetrendedFig}. The simple aperture photometric light curve has a scatter of 2527\,ppm, which improves to 685\,ppm after \texttt{everest} processing.

We searched the flattened data for periodic transit signals using the transit least squares algorithm (\texttt{TLS}) \citep{hippke2019a}, and found a prominent peak around 3.14 days as can be seen in \autoref{fig:transitandperiodgram}. We assessed the presence of additional candidate signals after modeling out the 3.14-d signal by re-running TLS, but did not find any with a significant signal detection efficiency (i.e., SDE$>$10). 

\subsection{Candidate Vetting with SPECULOOS Telescopes}
We followed up on the planetary candidate by observing with SPECULOOS Southern Observatory (SSO) two transit windows on UT 25 February 2020 by Ganymede and on UT 18 March 2020 by Io, and one transit window with SPECULOOS Northern Observatory on UT 18 May 2020 by Artemis. SSO is composed of four telescopes: Io, Europa, Ganymede and Callisto, which are installed at ESO Paranal Observatory (Chile) and have been operational since January 2018. SNO is currently composed of one telescope (Artemis), which is located at the Teide Observatory (Canary Islands, Spain) and operational since June 2019. All SPECULOOS telescopes are identical robotic Ritchey-Chretien (F/8) telescopes with an aperture of 1-m. They are equipped with Andor iKon-L cameras with e2v 2K $\times$ 2K deep-depletion CCDs, which provide a Field of View (FoV) of $12\,\arcmin\times12\,\arcmin$ and the corresponding pixel scale is $0.35\,\,\arcsec\,\mathrm{pixel^{-1}}$ \citep{delrez2018, jehin2018}.  To schedule those windows we used the SPeculoos Observatory sChedule maKer ({\fontfamily{pcr}\selectfont  SPOCK}), described in \cite{Sebastian2020}. Observations were made with an exposure time of 40\,s in an I+z filter, a custom filter (transmittance $>$90\% from 750\,nm to beyond 1000\,nm) designed for the observation of faint red targets usually observed by the SPECULOOS survey \citep{delrez2018,murray2020}. SSO data were then processed using the SSO Pipeline, which accounts for the water vapor effects known to be significant for differential photometry of redder hosts with bluer comparison stars \citep{murray2020}. SNO data were processed using \texttt{prose}, a Python-based data reduction package, which generates light curves from raw images \citep{garcia2020}.  It creates a stacked image to extract the positions of the stars in the field, and uses the positions to perform aperture annulus photometry. A differential light curve is produced using a weighted light curve derived from the field stars, while the instrumental systematics, such as pointing shift and full width half maximum, are recorded to assist later in detrending. We show detrended light curves from SPECULOOS in \autoref{fig:GNDObservation}, where we recovered the transit events within 1$\sigma$ of the calculated ephemeris from \textit{K2} data. Since these observations were obtained two years after \textit{K2} Campaign 15, they improve the precision of the transit ephemeris by an order of magnitude.  

\section{Analysis and Validation}
\label{sec:analysis}

\subsection{Stellar Host Characterization}
\subsubsection{Semi-empirical Stellar Parameters}
\label{sec:SED}

\begin{figure}[ht!]
\centering
\includegraphics[width=0.45\textwidth ]{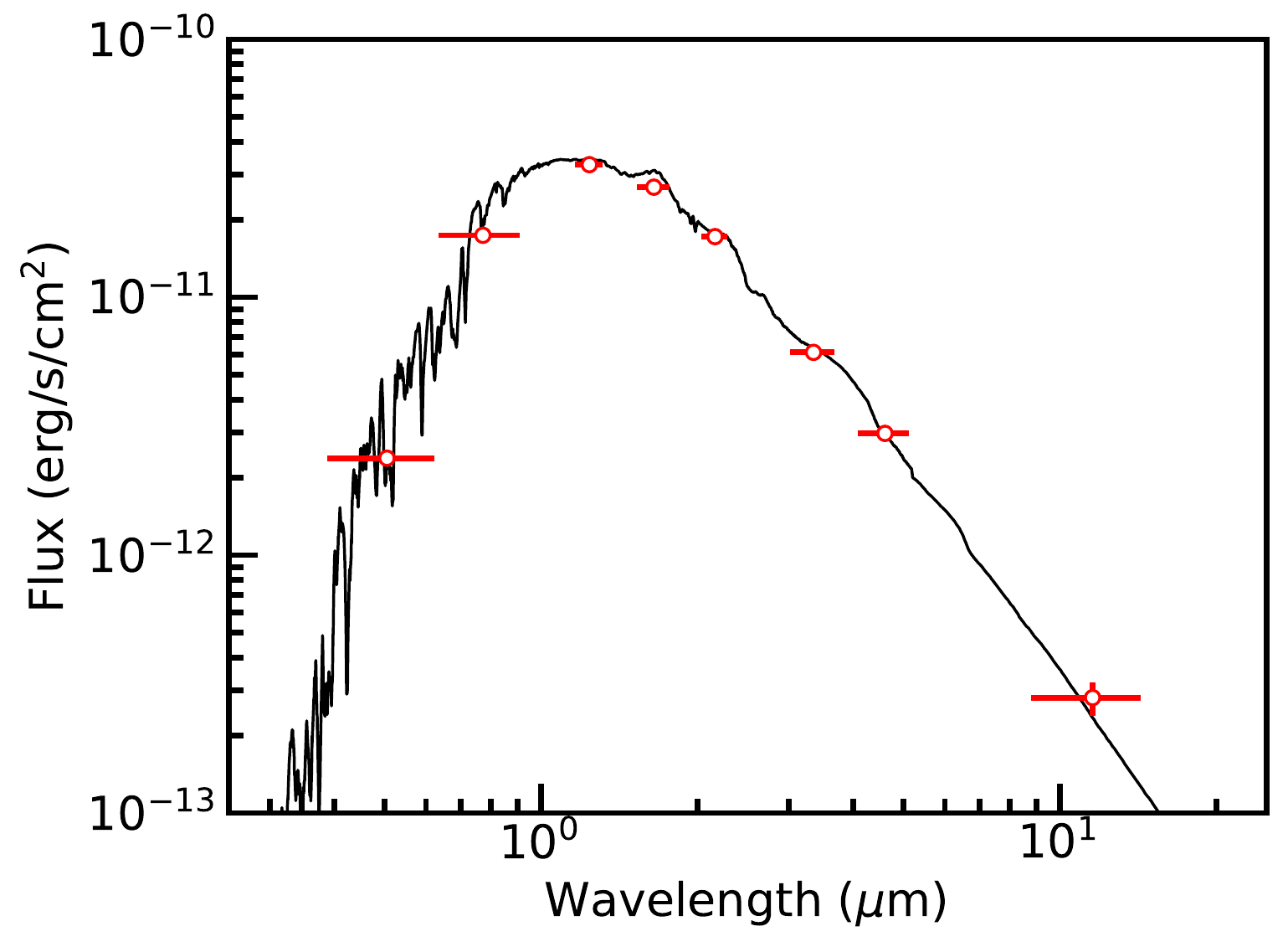} 
\figcaption{Spectral energy distribution of the host star.
Photometric fluxes used in the stellar characterization analysis (i.e., $G_{BP}$, $G_{RP}$, $J$, $H$, $K_{s}$, $W1$, $W2$, and $W3$) are shown as points with x-errors illustrating the filter bandpasses. Flux uncertainties are generally smaller than the marker size. For comparison, a BT-Settl model spectrum \citep{allard2012} is shown (black line) for a star with $T_\mathrm{eff} = 3300$\,K, $\log g = 5.0$, and [Fe/H] = 0.0, parameters similar to those derived from our SED analysis (Section~\ref{sec:SED}).
\label{fig:StellarSEDFit}}
\end{figure}

We constructed the spectral energy distribution (SED) of EPIC~249631677 using photometric magnitudes from \textit{Gaia} ($G_{BP}$ and $G_{RP}$; \citealt{GaiaDR2}) and the AllWISE source catalog ($J$, $H$, $K_{s}$, $W1$, $W2$, and $W3$; \citealt{cutri2013}). The corresponding fluxes for these magnitudes are tabulated on VizieR \citep{VizieR} and shown in \autoref{fig:StellarSEDFit} and Table~\ref{tab:stellar_props}. The parallax of EPIC~249631677 is $\pi = 17.61 \pm 0.09$\,mas,  which yields a distance of 56.8 $\pm$ 0.3\,pc \citep{GaiaDR2,stassun2018}.
We then derived the stellar luminosity $L_*$ by integrating over the SED, which yielded  $L_*=0.0041 \pm 0.0001 L_{\odot}$.

Two independent methods were applied to obtain stellar mass. First, we used the empirical $M_* - M_{K_s}$ relation (applying the metallicity obtained in Section~\ref{sec:spectro}) from \citet{2019ApJ...871...63M} to obtain $M_*=0.1721\pm0.0044 M_{\odot}$. We also applied stellar evolution modeling, using the models presented in \citet{2019ApJ...879...94F}, using as a constraint the luminosity inferred above and the metallicity derived in Sect.~\ref{sec:spectro}). We considered the stellar age to be $>$1 Gyr in the absence of signs of youth, such as presence of prominent flares \citep[][see Section~\ref{variability}]{ilin2019}. We obtained with this method $M_*=0.176 \pm0.004$ $M_{\odot}$. This uncertainty reflects the error propagation on the stellar luminosity and metallicity, but also the uncertainty associated with the input physics of the stellar models. We combined these two mass estimates as in \citet{2018ApJ...853...30V} to obtain  $M_*=0.174 \pm 0.004 M_{\odot}$ as our best estimate for the stellar mass of EPIC~249631677. Given its proximity, we expect minimal extinction (A$_v$) for the target; the SED fitting analysis described in Section~\ref{sec:binarity} similarly constrains it to be $0.02$ at $3\sigma$ confidence and we adopt that upper limit here. 
Finally, we note that given its luminosity, mass, and {\em Gaia} colors this star is likely to be fully convective  \citep{jao:2018,rabus:2019}.

\begin{figure*}[btp]
\centering
\includegraphics[width = 0.95\textwidth]{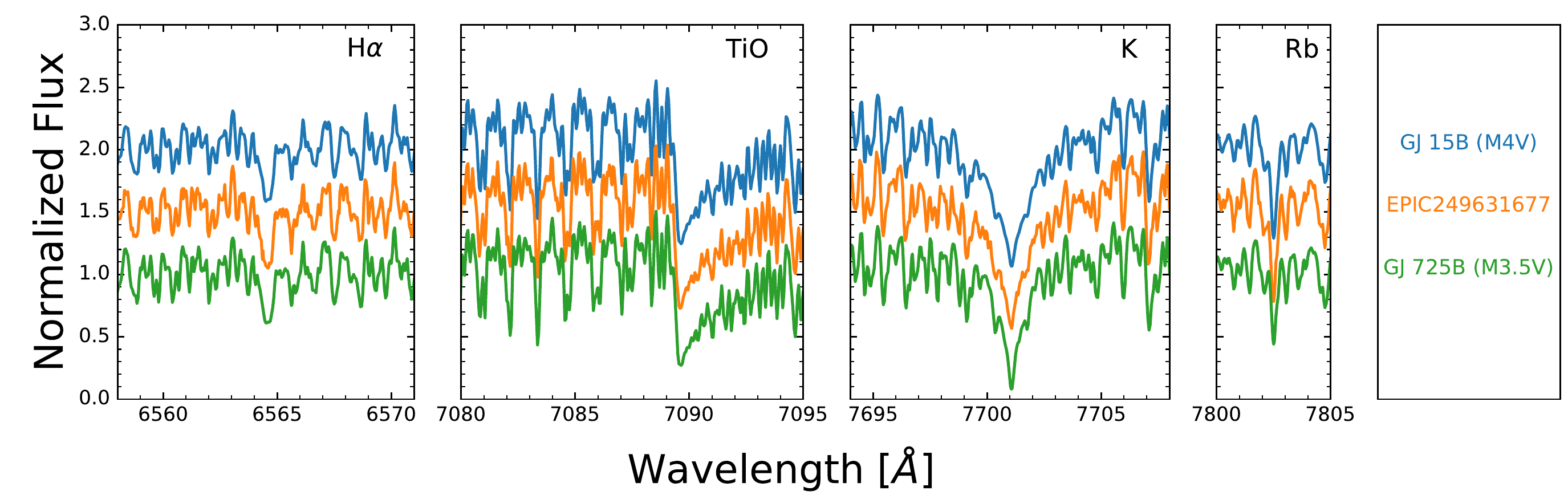}
\caption{Comparison of Keck/HIRES spectra of EPIC~249631677 (orange) with GJ 725B (green) and GJ~15B (blue) in the vicinity of the expected locations of H$\alpha$, TiO bands, K~I (7701.0\AA), and  Rb~I (7802.4\AA).  No secondary spectral lines, emission lines, or   rotational broadening are detected.
\label{fig:hires_spectra}}
\end{figure*}

Due to the absence of a strong constraint on the stellar density from the transits, we further obtained stellar radius,  surface gravity, effective temperature and density  from our evolutionary models. \autoref{tab:stellar_props} summarizes the results from this analysis, along with other properties of the star. Our values are consistent with those listed in the TESS Input Catalog \citep{TICv8}, and we adopt them for the remainder of this analysis.

\subsubsection{Reconaissance Spectroscopy}
\label{sec:spectro}
To confirm EPIC~249631677's stellar properties and better characterize the system, we acquired an optical spectrum using Keck/HIRES \citep{vogt:1994} on UT 30 May 2020. The observation took place in $0.6\arcsec$ effective seeing and using the C2 decker without the HIRES iodine gas cell, giving an effective resolution of $\lambda/\Delta \lambda \approx 55,000$ from 3600\,\AA\ to 7990\,\AA. We exposed for 1800~s and obtained SNR of roughly 23 per pixel. Data reduction followed the standard approach of the California Planet Search consortium \citep{howard:2010b}.

We used our Keck/HIRES radial velocity and Gaia DR2 data to estimate
the 3D galactic ($UVW$) space velocity using the online kinematics
calculator\footnote{\url{http://kinematics.bdnyc.org/query}} of
\cite{rodriguez:2016}. Following \cite{chubak:2012}, our Keck/HIRES spectrum gives a barycentric radial velocity of
$6.25\pm0.17$~km~s$^{-1}$. With the Gaia-derived coordinates, proper motion, and distance listed in Table~\ref{tab:stellar_props}, we find $(U,V,W)$ values of $(-17.02, -9.06, +33.66)$~km~s$^{-1}$, indicating a likely membership in the Milky Way's thin disk \citep{bensby:2014}.

Using the \texttt{SpecMatch-Empirical} algorithm \citep{yee:2017}, we
derive from our HIRES spectrum stellar parameters of
$T_\mathrm{eff}=3195\pm70$~K, $R_*=0.23\pm0.10R_\odot$, and
[Fe/H]$=-0.24\pm0.09$, consistent with the values tabulated in
Table~\ref{tab:stellar_props}.  The three best-matching stars in the
\texttt{SpecMatch-Empirical} template library are GJ~15B, GJ~447, and GJ~725B, 
which have spectral types of M3.5V, M4V, and M3.5V, respectively. Given the close match between the spectra of these stars and our target (see Fig.~\ref{fig:hires_spectra}), we therefore classify EPIC~249631677 as an M dwarf with subclass 3.5$\pm$0.5. We see no evidence of emission line cores at H$\alpha$, consistent with our determination that our target is not a young star. We see no evidence of spectral broadening compared to these three stars
\citep[which all have $v \sin i < 2.5$~km~s$^{-1}$;][]{reiners:2012}, so we set an upper limit on EPIC~249631677's projected rotational velocity of $<5$~km~s$^{-1}$, comparable to the spectral resolution of HIRES.

\subsubsection{Stellar Variability}
\label{variability}
The long-term variations apparent in the \texttt{everest} light curve (\autoref{fig:K2DetrendedFig}) are not evident in light curves from other reduction pipelines (e.g., \texttt{K2SFF}). These variations likely arise from systematics and are not reliable for estimating the stellar rotation period \citep{esselstein2018}. Similarly, no flares are apparent either in the \textit{K2} or SPECULOOS data. Flare rates peak for $\sim$M3.5 stars in TESS data \citep{guenther2020}. However, given the long integration time of 29.4\,minutes as well as a need for data processing which corrects for the saw-tooth pattern, flare signals, unless very prominent,  are expected to be difficult to detect in \textit{K2} long cadence data.

\begin{deluxetable}{lcc}
\tabletypesize{\footnotesize}
\tablewidth{0pt}
\tablecaption{Stellar properties. \label{tab:stellar_props}}
\tablehead{
    \colhead{Property} &
    \colhead{Value} &
    \colhead{Source}
}
\startdata
\multicolumn{3}{l}{\it Catalog names}\\
\hline
K2 ID  & 315                      & 1 \\
EPIC ID  & 249631677              & 2 \\
TIC ID   & 70298662               & 3 \\
2MASS ID & J15120519-2006307      & 4 \\
Gaia DR2 ID & 6255978483510095488 & 5 \\
\hline
\multicolumn{3}{l}{\it Astrometric Properties}\\
\hline
RA (J2000, hh:mm:ss)                & 15:12:05.19      & 5 \\
Dec (J2000, dd:mm:ss)               & -20:06:30.55     & 5 \\
Distance (pc)                       & {56.8} $\pm$ 0.3 & 5 \\
$\mu_\mathrm{RA}$ (mas\,yr$^{-1}$)  & -120.3 $\pm$ 0.2 & 5 \\
$\mu_\mathrm{Dec}$ (mas\,yr$^{-1}$) &   74.7 $\pm$ 0.1 & 5 \\
Barycentric Radial Velocity (km~s$^{-1}$) & +6.25 $\pm$ 0.17 & 7 \\  
\hline
\multicolumn{3}{l}{\it Photometric Properties}\\
\hline
$B$ (mag)       & 18.656 $\pm$ 0.162 & 3 \\
$V$ (mag)       & 17.67 $\pm$ 0.2    & 3 \\
$G_{BP}$ (mag)       & 17.3648 $\pm$ 0.0134 & 5 \\
$G$ (mag)       & 15.6791 $\pm$ 0.0010 & 5 \\
$G_{RP}$ (mag)       & 14.4183 $\pm$ 0.0028 & 5 \\
$J$ (mag)       & 12.665$\pm$ 0.022  & 6 \\
$H$ (mag)       & 12.134 $\pm$ 0.027 & 6 \\ 
$K_{s}$ (mag)   & 11.838$\pm$ 0.023  & 6 \\
WISE 3.4 (mag)  & 11.631$\pm$0.024   & 6 \\
WISE 4.6 (mag)  & 11.436$\pm$0.023   & 6 \\
WISE 12.0 (mag) & 11.068$\pm$0.156   & 6 \\
\hline
\multicolumn{3}{l}{\it Derived Fundamental Properties}\\
\hline
Mass, $M_{*}$ ($M_{\sun}$)                  & 0.174 $\pm$ 0.004 & 7 \\
Radius, $R_{*}$ ($R_{\sun}$)                & 0.196   $\pm$   0.006  & 7 \\
Density, $\rho_{*}$ (g\,cm$^{-3}$)          & 32.6  $\pm$   1.0    & 7 \\
Luminosity, $L_{*}$ ($L_{\sun}$)            & 0.0041 $\pm$ 0.0001 & 7 \\
Effective Temperature, $T_\mathrm{eff}$ (K) & 3300   $\pm$ 30     & 7 \\
Surface Gravity, $\log g$ (cgs)             & 5.094  $\pm$ 0.006  & 7 \\
Age (Gyr)                                   & $>1$                & 7, 9 \\
Metallicity, [Fe/H]                         & -0.24  $\pm$ 0.09   & 8 \\  
Spectral Type                               & M(3.5$\pm$0.5)V     & 8 \\  
Projected Rotation, $v \sin i$ (km~s$^{-1}$) & $<5$             & 8 \\  
Extinction, $A_{V}$                         & $<0.02$             & 9 \\
\enddata
\tablerefs{
(1) This work.
(2) \citet{epic}.
(3) \citet{TICv8}.
(4) \citet{cutri2003}.
(5) \citet{GaiaDR2}.
(6) \citet{cutri2013}.
(7) This work, evolutionary model analysis.
(8) This work, Keck/HIRES analysis.
(9) This work, SED analysis.
}
\end{deluxetable}

\subsection{Vetting}
\begin{figure*}[ht!]
\centering
\includegraphics[trim={2cm 0 4cm 1cm},clip, width=\textwidth ]{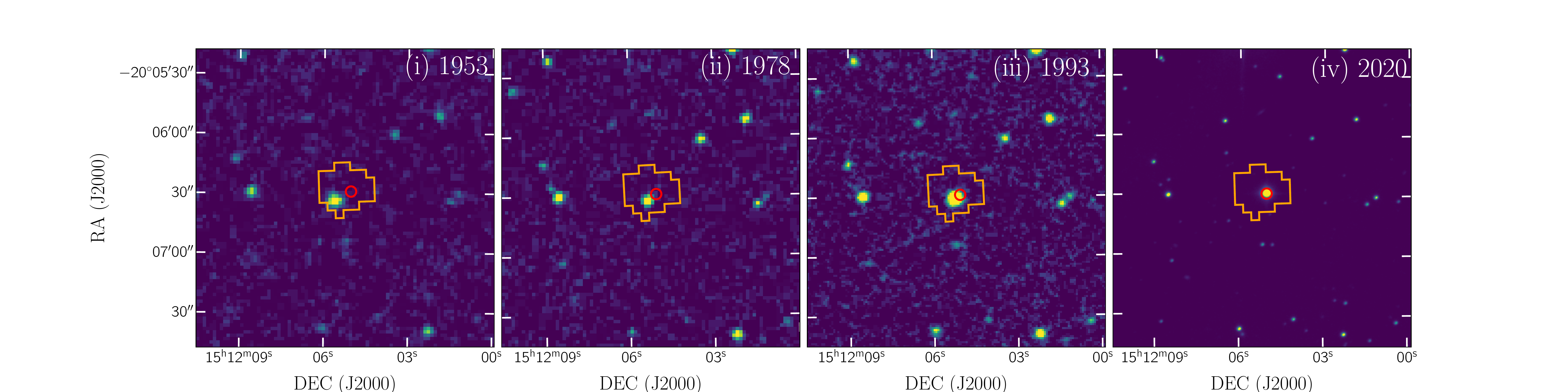} 
\figcaption{Set of archival image used to check for background objects. The orange polygon represents the aperture used in K2 by \texttt{everest} pipeline, while the red circle represents the aperture used for extracting photometry using SNO. \textbf{i)} POSS I Survey image from 1953 does not shown any bright object in the current SSO aperture. \textbf{ii)} Image from Hubble Guide Star Catalogue 1  from 1978. \textbf{iii)} Image from POSS II from 1993. \textbf{iv)} Median stacked image from Artemis, SNO observation made on May 18 2020. All three archival figures do not show any background object at the current position of EPIC 249631677. \label{fig:ArxivalImage}}
\end{figure*}
In order to produce a transit depth on the level of 0.2\% in the light curve of the primary target, a background eclipsing binary producing eclipses with depths of 25\% to 50\% would have to be 5.25 to 6.0 mag fainter than the target, respectively. Qualitatively, the odds of EPIC~249631677 hosting a planet are higher than the odds of such magnitude contrast eclipsing binary being present within the SPECULOOS aperture, given occurrence rates of M-dwarf planets \citep{dressing2013, mulders2015, ullman2019}. A stringent quantitative constraint can be placed using the ingress/egress duration ($T_{12}/T_{34}$) compared to the total transit duration ($T_{14}$) \citep[Equation 21]{seager2003}. Such a test yields an upper limit on the relative radius of the transiting body. By assuming equal effective surface temperatures, the lower magnitude limit $\Delta m$ (corresponding to a flux difference $\Delta$F) for a blended binary mimicking a signal of depth $\delta$ is given by:

\begin{equation}
\begin{split}
    \Delta F &= \left(\frac{1 - T_{23}/T_{14}}{1 + T_{23}/T_{14}}\right)^2 \\
    \implies \Delta m  &= 2.5 \log_{10}\left(\frac{\Delta F}{\delta}\right). 
\end{split}    
\end{equation}

Using the posterior for the transit fit (See Section~\ref{sec:transitfitting}), we find for EPIC~249631677 that such a background object can be fainter at most by 1.73 mag at the 3$\sigma$ level. Fortunately, EPIC~249631677 has a significant proper motion, $\sim$140\,mas yr$^{-1}$, which allows us to investigate the presence of background sources at its current sky position. We looked at archival imaging of EPIC 246331677 going back to 1953\footnote{\href{http://stdatu.stsci.edu/cgi-bin/dss\_form}{http://stdatu.stsci.edu/cgi-bin/dss\_form}}. A POSS I plate from 1953 is the publicly available oldest image of EPIC 249631677, and it does not show any background source at the current position of the target as shown in \autoref{fig:ArxivalImage}. The plate is sensitive to objects at least 3.5 magnitudes fainter than the target. Similarly, the Hubble Guide Star Catalogue (GSC), with a limiting magnitude of 20 \citep{lasker1990}, does not show any background source.  While POSS II would go the deepest in terms of limiting magnitude \citep[20.8,][]{reid1991}, the star has moved appreciably closer to its current location, precluding a definitive measurement from this image. Overall, using archival images we can rule out the possibility of the transit signal originating from background star at a high level of confidence.

\subsubsection{Binarity of the Host Star}
\label{sec:binarity}
Despite the lack of background sources, the host star could produce a false-positive transit signal if it were a grazing eclipsing binary or a hierarchical eclipsing binary.
We investigated the evidence for host star binarity using the \texttt{isochrones} software package \citep{isochrones}, which performs isochrone fitting in the context of the MESA \citep{MESA2011, MESA2013, MESA2015} Isochrones and Stellar Tracks database \citep{MIST0, MIST1}.
Single-star and binary evolutionary models are available within \texttt{isochrones}, and the inference is performed via the nested sampling algorithm MULTINEST \citep{MULTINEST} (as implemented in the \texttt{PyMultiNest} software package \citep{PyMultiNest}), which allows for direct comparisons of the Bayesian evidence $\ln Z$.

We tested both single-star and binary models using the priors on photometric magnitudes and stellar distance described in Section~\ref{sec:SED}. The inferred properties from the single-star model fit are consistent with those given in \autoref{tab:stellar_props} at the 2$\sigma$ level.
The $\ln Z$ for the single-star model is $-213.86 \pm 0.04$, whereas the $\ln Z$ for the binary model is $-229.6 \pm 0.2$. According to \citet{kass95}, the corresponding Bayes factor of 16 indicates ``decisive'' evidence in favor of the single-star model.

We also examined our Keck/HIRES spectrum for secondary lines that would indicate
the presence of another star following the approach of \cite{kolbl:2015}. We found no evidence of additional lines down to the method's standard sensitivity limit of $\Delta V=5$~mag for
$\Delta v > 10$~km~s$^{-1}$, consistent with EPIC~249631677 being a single, isolated star.
We therefore conclude that the available data strongly support EPIC~246331677 being a single star.

\subsubsection{Photometric Tests}
We performed a series of tests on the photometric data to rule out false-positive scenarios. First, we performed an even-odd test on the target using \textit{K2} photometry. The even and odd transits are consistent with one another in transit depth to within 1$\sigma$. We also looked for secondary eclipses in the phase-folded light curve and found none to be present. Note that since we observe consistent signals in both the \textit{K2} and SPECULOOS data sets, we can rule out the signal originating from systematics. The transit depth in SPECULOOS observations with I+z filter, which is redder than Kepler bandpass, are consistent to K2 transit depths within 1$\sigma$ level, keeping up with the expectation of the achromatic nature of planetary transit. Furthermore, a massive companion, such as a faint white dwarf, can be ruled out using the ellipsoidal variation, which puts a 3$\sigma$ upper limit on the mass of any companions at the given orbital period of the transit signal as $\sim$100\,$M_{\rm Jup}$ \citep{morris1985, niraula2018}. From the transit fit, we can rule out a grazing eclipse originating from a larger transiting object (i.e. $\ge$ 2R$_{\oplus}$) at ${>}3\sigma$ confidence. Together, these tests rule out the object at 3.14 days being a massive companion.

\subsection{Transit Fitting}
\label{sec:transitfitting}
We used the refined estimates of the host properties together with a joint analysis of the \textit{K2} and SPECULOOS light curves to derive the planetary properties. In order to calculate the transit model, we used \texttt{batman} \citep{kreidberg2015}. We simultaneously model both the \textit{K2} observation as well as the ground-based observations with 21 parameters in a Monte Carlo Markov Chain (MCMC) framework using the \texttt{emcee} package \citep{mackey2013}. We use a Gaussian prior on the scaled semi-major axis of the orbit $a/R_{*}$ of $\mathcal{N}$(25.72, 0.27), derived using Eq.~30 from \citet{winn2010} along with the stellar density of 32.6 $\pm$ 1.0 g\,cm$^{-3}$ (see Section~\ref{sec:SED}) and orbital period of 3.1443 days from the TLS search. As for the limb darkening, we use the non-informative $q_{1}$, $q_{2}$ parameterization of the quadratic limb-darkening law as suggested by \citet{kipping2013}. We fixed the eccentricity to 0, given that the expected time of circularization is 50 Myr \citep[assuming a quality factor $Q_p\sim500$,][]{goldreich1966, patra2017}, which is at least an order of magnitude smaller than the estimated age of the system.  For \textit{K2} data, we supersample the transits by a factor of 15 in \texttt{batman} to take into account the effect of non-negligible integration time. As for the ground-based data, we use second-order polynomials to detrend against the observables airmass and FWHM.  

We ran the MCMC for 50,000 steps with 150 walkers performing a combined fit of \textit{K2} and SPECULOOS data, and use the last half of the run to build the parameter posteriors. We assessed the convergence of walkers using the suggested autocorrelation test for \texttt{emcee}. The resulting median values from the fit with $1\sigma$ deviation are reported in \autoref{table:transitparams}, while the best-fit transit models are shown in \autoref{fig:transitandperiodgram} and \autoref{fig:GNDObservation}.  

\begin{deluxetable}{lc}
\tablecaption{\label{table:transitparams}Transit Fit Parameters}
\tablehead{
\colhead{Property} &
\colhead{Value} 
}
\startdata
Period (Days) & 3.1443189 $\pm$ 0.0000049\\
$T_{0}- 2450000$ (BJD) & 7990.8620$^{+0.0010}_{-0.0011}$\\
$R_{p}/R_{_*}$ & 0.0444 $\pm$ 0.0024\\
Radius $(\ER)$ & 0.950 $\pm$ 0.058 \\
a/R$_*$  & 25.72$^{+0.16}_{-0.17}$ \\
Inclination (Deg)  &  88.74 $^{+ 0.21}_{- 0.16}$ \\
$b$  & 0.565$^{+0.070}_{-0.092}$ \\
$u_1$ (\textit{Kepler}) & 0.80 $^{+ 0.57}_{-0.53}$\\
$u_2$ (\textit{Kepler}) & -0.12$^{+ 0.49}_{- 0.44}$\\
$u_1$ (\textit{\rm I+z})& 0.49$^{+0.54}_{-0.35}$\\
$u_2$ (\textit{\rm I+z})& 0.06$^{+ 0.42}_{- 0.39}$\\
$T_{14}$ (Hours) & 0.821$^{+ 0.047}_{-0.043}$\\
Instellation (S$_\oplus$) & 7.45$_{-0.44}^{+0.48}$\\
$T_{\rm eq}^{\dagger}$(K) & 460 $\pm$ 5\\
\enddata
\tablenotetext{\dagger}{Calculated assuming Bond albedo of 0.}
\end{deluxetable}

\begin{figure*}[ht!]
\centering
\includegraphics[trim={0cm 0cm 2cm 1cm},clip, width=0.95\textwidth ]{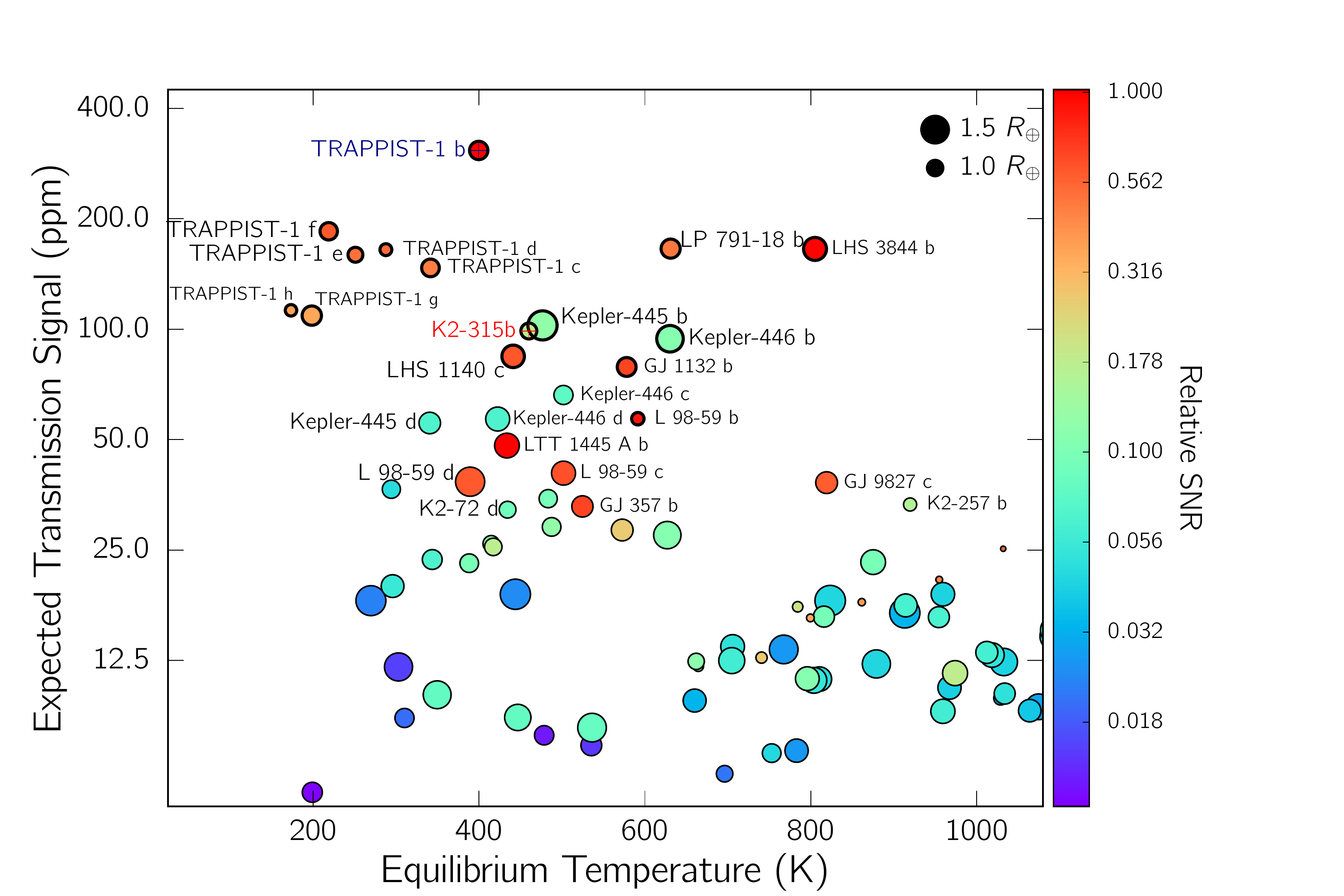} 
\figcaption{Most promising terrestrial planets for atmospheric characterization. Point colors illustrate the SNR of a \textit{JWST/NIRSPEC} observation relative to TRAPPIST-1~b.  SNR below 1/100th of TRAPPIST-1b and transmission signal less than 5 ppm have been removed to enhance readability of the figure. The planets for which the presence of an atmosphere could be assessed by \textit{JWST} within $\sim$ 50 transits are encircled in black, if their atmospheric signals are above \textit{JWST}'s threshold of $\sim$50 ppm. The rest of the uncircled pool of terrestrial planets may be accessible with the successors of \textit{JWST} if ten times better performance can be achieved. The size of the circle is proportional to the size of the planet. Circles for 1.5\,R$_\oplus$ and 1.0\,$R_\oplus$ are drawn in the upper right corner for reference. \label{fig:AtmosphericSignal}}
\end{figure*}

\section{Future Prospects}
\label{sec:future}
The search for transiting planets around small stars has been motivated in large part by their potential for atmospheric characterization. Owing to the size and proximity of its host, K2-315b is thus one of the few known terrestrial exoplanets possibly amenable for atmospheric characterization in the next two decades. In order to quantify and contextualize its prospects for atmospheric study, we followed the same approach as for TRAPPIST-1 in \citet{gillon2016} \citep[see][]{dewit2013}, focusing here on all known terrestrial planets. We selected terrestrial planets as planets with a reported radius below 1.6\,R$_{\oplus}$ in the NASA Exoplanet Archive\footnote{\href{https://exoplanetarchive.ipac.caltech.edu}{https://exoplanetarchive.ipac.caltech.edu}} \citep{rogers2015,fulton2017}.  
We thus derive the amplitude of the planets' signals in transmission as:
\begin{equation}
\begin{split}
S &=\frac{2 R_p h_{\rm eff}}{R_*^2}, {\rm with}\\
h_{\rm eff} &= \frac{7 k T}{\mu g},
\label{eqn:AtmS}
\end{split}
\end{equation}
where $R_p$ is the planetary radius, $R_*$ is the stellar radius, and $h_{\rm eff}$ is the effective atmospheric height, $\mu$ is the atmospheric mean molecular mass, $T$ is the atmospheric temperature and $g$ is the local gravity. We assume $h_{\rm eff}$ to cover seven atmospheric scale heights, $\mu$ the atmospheric mean molecular mass to be 20 amu, and the atmospheric temperature to be the equilibrium temperature for a Bond albedo of 0. For the planets with missing masses, we estimated $g$ using the model of \citet{chen2017}. 

The signal amplitudes are reported in \autoref{fig:AtmosphericSignal} together with the SNR relative to TRAPPIST-1~b's, calculated by scaling the signal amplitude with the hosts' brightness in $J$ band. We find that K2-315b fares closely to the outer planets of TRAPPIST-1 in terms of potential for atmospheric exploration with \textit{JWST}---its warmer and thus larger atmosphere compensating for its larger star. In fact, its relative SNR for transmission spectroscopy is half those of TRAPPIST-1~f--h, meaning that assessing the presence of a $\mu\sim20$ atmosphere around the planet would require of the order of 40 transits---four times the $\sim$10 transits required for a similar assessment for TRAPPIST-1~f--h \citep{Lustig2019}. K2-315b is thus at the very edge of \textit{JWST}'s capability for atmospheric characterization, mostly due to its ``large'' host star. The derivation above allows us to rank planets in terms of relative potential for atmospheric characterization, assuming a similar atmospheric scenario. In practice, a significant difference in atmospheric mean molecular mass, surface pressure, and/or cloud/haze altitude will strongly affect the actual potential of a planet for characterization \citep{Lustig2019}. For example, in some cases clouds could render the characterization of the favorable TRAPPIST-1~e difficult even with the generation of instruments following the JWST \citep[i.e., LUVOIR,][]{pidhorodetska2020}.

With an estimated radial velocity semi-amplitude of 1.3~m~s$^{-1}$ (assuming a mass comparable to that of Earth),  the planet could be  accessible for mass measurements using modern ultra-precise radial velocity instruments. Such possibilities and a ranking amongst the 10 best-suited Earth-sized planets for atmospheric study, EPIC 249631677 b will therefore play an important role in the upcoming era of comparative exoplanetology for terrestrial worlds. It will surely be a prime target for the generation of observatories to follow \textit{JWST} and bring the field fully into this new era.

\vspace{0.5cm}

{\bf Acknowledgments:}
This project makes use of publicly available \textit{K2} data. P.N. would like to acknowledge funding for Kerr Fellowship and Elliot Fellowship at MIT. J.d.W. and MIT gratefully acknowledge financial support from the Heising-Simons Foundation, Dr. and Mrs. Colin Masson and Dr. Peter A. Gilman for Artemis, the first telescope of the SPECULOOS network situated in Tenerife, Spain. B.V.R. thanks the Heising-Simons Foundation for support. V.V.G. is a F.R.S.-FNRS Research Associate. M.G. and E.J. are F.R.S.-FNRS Senior Research Associates.  This work was also partially supported by a grant from the Simons Foundation (PI Queloz, grant number 327127), and funding from the European Research Council under the European Union's Seventh Framework Programme (FP/2007-2013) ERC Grant Agreement Number 336480, from the ARC grant for Concerted Research Actions financed by the Wallonia-Brussels Federation, from the Balzan Prize Foundation, from F.R.S-FNRS (Research Project ID T010920F), from the European Research Council (ERC) under the European Union’s Horizon 2020 research and innovation programme (grant agreement n$^\circ$ 803193/BEBOP), and from a Leverhulme Trust Research Project Grant n$^\circ$ RPG-2018-418.

\facilities{Keck:I (HIRES), Kepler, MAST (HLSP, K2), SNO, SSO}

\software{ \tt{astropy} \citep{astropy1, astropy2}, \tt{batman} \citep{kreidberg2015}, \texttt{emcee} \citep{mackey2013},  \texttt{isochrones} \citep{isochrones}, \texttt{prose} \citep{garcia2020}, \texttt{SPOCK} \citep{Sebastian2020}, \texttt{transitleastsquares} \citep{hippke2019a}, \texttt{w{\={o}}tan} \citep{hippke2019b}}

\bibliographystyle{aasjournal}
\bibliography{Bibliography.bib}{}

\end{document}